\documentclass[a4paper,11pt]{article}
\usepackage{amsmath}
 
\usepackage{amsfonts}
\usepackage{tabularx,ragged2e,booktabs,caption}
\usepackage{setspace}
\usepackage{amssymb}
\usepackage{color}
\usepackage[round,authoryear]{natbib}
\usepackage{graphicx}
\usepackage{graphics}
\usepackage{lineno}
\usepackage{epstopdf}
\usepackage{lipsum}
\usepackage[T1]{fontenc}
\usepackage[utf8]{inputenc}
\usepackage{authblk}
\usepackage{epstopdf}
\usepackage{enumitem}
\usepackage{appendix}
\usepackage{rotating}
\usepackage{array}
\newcolumntype{L}[1]{>{\raggedright\arraybackslash}p{#1}}
\newcolumntype{C}[1]{>{\centering\arraybackslash}p{#1}}
\newcolumntype{R}[1]{>{\raggedleft\arraybackslash}p{#1}}
\usepackage{geometry}
\usepackage{booktabs}
\usepackage{tabularx, array}
\usepackage{dcolumn}
\usepackage{xcolor,colortbl}
\definecolor{Gray}{gray}{0.9}
\newcolumntype{a}{>{\columncolor{Gray}}c}
\usepackage[flushleft]{threeparttable}
\usepackage{caption}
\usepackage{hhline}

\begin{document}
\title{Critical Decisions for Asset Allocation via Penalized Quantile Regression}

\author{Giovanni Bonaccolto\thanks{School of Economics and Law, University of Enna 
``Kore'', Italy. Email: giovanni.bonaccolto@unikore.it} }
\date{This version: \today}
\maketitle


\begin{abstract}
We extend the analysis of investment strategies derived from penalized quantile regression models, introducing alternative approaches to improve state\textendash of\textendash art asset allocation rules. First, we use a post\textendash penalization procedure to deal with overshrinking and concentration issues. Second, we investigate whether and to what extent the performance changes when moving from convex to nonconvex penalty functions. Third, we compare different methods to select the optimal tuning parameter which controls the intensity of the penalization. Empirical analyses on real\textendash world data 
show that these alternative methods outperform the simple LASSO. This evidence becomes stronger when focusing on the extreme risk, which is strictly linked to the quantile regression method.  
\end{abstract}


\section{Introduction}\label{sec:intro}

Recent financial crises have shown the impact of extreme losses on the overall system, highlighting the need to design better investment strategies and risk management policies to minimize or predict more accurately the effects of tail events. Therefore, the focus has begun to shift toward pessimistic asset allocation strategies, designed to minimize measures of extreme risk, going beyond the classical mean\textendash variance theory pioneered by \cite{Ma52}. Among the various measures of extreme risk studied in the theoretical and empirical literature, the expected shortfall (ES) introduced by the seminal works of \cite{RoUr00} and \cite{AcTa02} has received a relevant attention (see, among many others, \cite{Yamai2005}, \cite{Ch08}, \cite{Assa2015}, \cite{doi:10.1080/1350486X.2018.1492347}, \cite{Kratz2018} and \cite{PATTON2019388}). This is mainly due to the fact that the ES has useful mathematical and statistical properties, being a coherent risk measure according to \cite{ArDeEbHe99}. 
 
Several works used the ES as a target measure in portfolio optimization problems (see, e.g., \cite{Gilli2002} and 
\cite{doi:10.1080/14697680701422089}). \cite{BaKoKo04} showed that it is possible to efficiently minimize the ES of a given portfolio using the quantile regression method introduced by \cite{KoBa78}. Indeed, the coefficients of a quantile regression model, whose variables are computed from the returns yielded by the selected stocks, coincide with the weights of the portfolio with minimum ES. 
 Financial portfolios are typically built using a large number of stocks to exploit the advantages arising from their diversification. On the other hand, the larger the portfolio dimensionality is, the larger the number of parameters to estimate. The resulting accumulation of estimation errors is then a critical issue, especially in terms of out\textendash of\textendash sample performance.  Moreover, financial returns are typically    
highly correlated and the resulting portfolio weights turn out to be poorly determined,
exhibiting a relevant variance. 

In a context in which the optimal weights of a given portfolio are derived from regression models, regularization techniques have proven to be an effective tool to deal with the curse of dimensionality. For instance, \cite{FaPaWi14} showed that sparse portfolios derived from penalized regression models provide improvements in terms of a lower concentration and turnover. \cite{GiuzioPaterlini2018} highlighted the positive impact of regularized models on the portfolio performance during stressed market conditions. \cite{Kremer2018} used regularization techniques to minimize the risk in multi\textendash factor portfolios. The studies cited above, as well as many others in the literature, used various penalty functions to regularize a given regression model. Among them, the $\ell_1$\textendash norm penalty, leading to the Least Absolute Shrinkage and Selection Operator (LASSO) introduced by \cite{Ti96}, is one of the most commonly used penalization method in the financial literature (see, e.g., \cite{BrDa09}, \cite{DMGaNoUp09}, \cite{FaZhYu12} and \cite{YeYe14}). In recent years, the LASSO has gained attention in the statistical literature not only when implemented on standard linear regressions but also on quantile regression models (see, e.g., \cite{Ko05}, \cite{LiZh08} and \cite{BeCh11}). In contrast, applications of penalized quantile regression models to build asset allocation strategies are still limited. 

In this study, we focus on \cite{BoCaPa}, who analyzed the performance of financial portfolios derived from quantile regression models including an $\ell_1$\textendash norm penalty, highlighting the resulting improvements in terms of out\textendash of\textendash sample performance. However, the work of \cite{BoCaPa} opened the door to further research, that we investigate in this study. An important research question is related to the use of the LASSO. Indeed, although the LASSO has many remarkable properties, it suffers from some limitations. For instance, it typically provides biased estimates, overshrinking the retained variables (see, e.g., \cite{Fan2001}). Under a financial viewpoint, this overshrinking causes a problematic consequence. Indeed, as highlighted by \cite{BoCaPa}, the asset having the role of response variable in the underlying regression model is typically overweighted. This raises problems in terms of portfolio concentration. In this study, we address this issue using a post\textendash LASSO procedure, that is a well\textendash known method in the financial econometrics literature. For instance, \cite{Hautsch14} used a post\textendash LASSO method to estimate pessimistic financial networks. Moreover, we also use nonconvex penalty functions, such as the Smoothly Clipped Absolute Deviation (SCAD) and the Minimax Concave Penalty (MCP), introduced, respectively, by \cite{Fan2001} and \cite{Zhang2010}, which have the oracle property as defined by \cite{Fan2001}. Many works in the literature used nonconvex penalty functions in standard linear regression models to build financial portfolios (see, among others, \cite{Giuzio}). Nevertheless, to the best of our knowledge, this is the first study in which a post\textendash LASSO procedure and nonconvex penalty functions are used within a quantile regression model to design asset allocation strategies.

Another relevant point is related to the method we use to control the intensity of the penalization; that is, the optimal tuning parameter. \cite{BoCaPa} used the data\textendash driven method that \cite{BeCh11} developed for quantile regression models including an $\ell_1$\textendash norm penalty. However, other selection methods are available in the literature. In this study, we also use a Bayesian Information Criterion (BIC) that \cite{Ryung2014} designed for quantile regression models and the $K$\textendash fold cross\textendash validation, that is commonly used in applied machine learning, being flexible and easy to understand and implement (see, e.g., \cite{HaTiFr09}). In contrast to the method of \cite{BeCh11}, the BIC and the $K$\textendash fold cross\textendash validation are flexible to be used on any penalized quantile regression model, regardless of the specification of its penalty function. To the best of our knowledge, the BIC of \cite{Ryung2014} and cross\textendash validation techniques have never been tested in building financial portfolios from quantile regression models.


We evaluate the competing penalty functions and regularization parameter selection methods, as well as the opportunity of using a post\textendash penalization procedure, by means of an extensive empirical analysis. The out\textendash of\textendash sample results highlight the significant improvements yielded by some of the resulting portfolios. Among them, we emphasize the outperformance of the rule in which we combine the post\textendash LASSO procedure with the selection method of \cite{BeCh11}. This evidence becomes stronger when focusing on the risk dimension, in terms of both standard deviation and expected shortfall of the out\textendash of\textendash sample portfolio returns. In particular, the latter acquires a central role in this study, as it represents the risk measure we minimize when estimating a quantile regression model. We also observe that nonconvex penalty functions often outperform the simple LASSO, providing then an additional valid alternative to state\textendash of\textendash art methods. Moreover, our  proposal turns out to be an effective tool to deal with the portfolio concentration issues highlighted by \cite{BoCaPa}, mitigating the weight of the stock whose return represents the response variable in the underlying quantile regression model.
Finally, our portfolio strategies are computationally efficient, providing then an advantageous tool for different financial agents, such as hedge funds, to improve the performance of their portfolios, while controlling the risk of incurring large losses during tail events.

The paper is structured as follows. Section \ref{sec:methods} describes the penalized quantile regression models we estimate to build financial portfolios. Section \ref{sec:empsetup} reports the details of the empirical set\textendash up. Section \ref{sec:empirical_results} reports and discusses the main empirical results, whereas Section \ref{sec:conclusions} concludes the paper.

\section{Methods}\label{sec:methods}

Let $\textbf{r}_t=\left[r_{1,t}\; r_{2,t}\; \cdots\; r_{N,t}\right]$ be an $1 \times N$ vector of returns yielded by $N$ assets that we observe at time $t$, for $t=1,...,T$.\footnote{To be precise, our dataset includes $Q>T$ weekly returns for each stock. However, we divide our time series into $Q-T$ equally\textendash sized subsamples, each of which includes $T$ weeks, to implement a rolling window procedure in the empirical analysis (see Section \ref{sec:empsetup} for additional details). The methods described in this section refer to a generic sample which includes the interval $[1,T]$; that is, the first subsample in the rolling window scheme. However, the theory described in this section equally
applies to the remaining subsamples spanning the intervals   
$[2,T+1]$,...,$[Q-T+1,Q]$.} We also define an $1 \times N$ vector of portfolio weights, denoted as $\textbf{w}=\left[w_1\; w_2\; \cdots\; w_N\right]$. We focus on portfolios satisfying the budget constraint, such that $\textbf{w} \left(\textbf{1}_N\right)^\prime = 1$, where $\textbf{1}_N$ is an $1 \times N$ unit vector. We then compute the portfolio return as $r_{p,t}=\textbf{r}_t \textbf{w}^\prime$, for $t=1,...,T$. 

Following \cite{FaZhYu12}, we can efficiently rewrite the return of a portfolio satisfying the budget constraint. For this purpose, we must select a reference index $s\in \{1,...,N\}$ and  then define the variables $x_{s,t}=r_{s,t}$ and $x_{j,t}=x_{s,t} - r_{j,t}$, for $1 \leq j \leq N$ and $j \neq s$. As a result, the return of this portfolio is defined as follows: 
\begin{equation}\label{eq:portret}
r_{p,t} = x_{s,t}-\sum_{\substack{j=1 \\ j\neq s}}^{N}w_j x_{j,t},
\end{equation}
where the weight of the $s$\textendash th asset is equal to $w_s=1-\sum_{\substack{j=1 \\ j\neq s}}^{N} w_j$ to satisfy the condition $\textbf{w} \left(\textbf{1}_N\right)^\prime = 1$.

\cite{BaKoKo04} introduced a method building on the quantile regression method \citep{KoBa78} to minimize the expected shortfall (ES) at the level $\tau$ of a given portfolio, where $\tau$ takes low values, typically in the interval $(0, 0.05]$ (see \cite{AcTa02}, \cite{RoUr00} and \cite{ArDeEbHe99} for a detailed description of the properties of the ES). \cite{BoCaPa} studied the performance of large portfolios keeping the focus on the pessimistic asset allocation of \cite{BaKoKo04}. The number of parameters to estimate increases with $N$. The accumulation of estimation errors becomes then a critical issue when $N$ takes large values. \cite{BoCaPa} dealt with the curse of dimensionality using the Least Absolute Shrinkage and Selection Operator (LASSO) introduced by \cite{Ti96}. In particular, by adding an $\ell_1$\textendash norm penalty to the objective function of the standard quantile regression model, \cite{BoCaPa} derived the portfolio weights from the following minimization problem:
\begin{equation} \label{lassoqr}
\operatorname*{arg\,min}_{(\textbf{w}_{(-s)},\;\mu)\in \mathbb{R}^N} \frac{1}{T}\sum_{t=1}^{T} \rho_\tau \left(x_{s,t}-\sum_{\substack{j=1 \\ j\neq s}}^{N}w_j x_{j,t}-\mu\right) + \lambda \frac{\sqrt{\tau (1-\tau)}}{T} \sum_{\substack{j=1 \\ j\neq s}}^{N} \widehat{\sigma}_j |w_j|,
\end{equation}
where $\rho_\tau(u)=u\left(\tau - \mathbb{I}_{\{u<0\}}\right)$ is the asymmetric loss function used by \cite{KoBa78}, $\mathbb{I}_{\{\cdot\}}$
is an indicator function which takes the value of one if the condition in $\{\cdot\}$ is true, the value of zero otherwise, $\textbf{w}_{(-s)}$ is the vector of the portfolio weights without $w_s$ (which is computed in a subsequent step as $w_s=1-\sum_{\substack{j=1 \\ j\neq s}}^{N} w_j$), $\widehat{\sigma}_j$ is the sample standard deviation of the variable $x_{j,t}$ and $\lambda>0$ is a tuning parameter. 

The statistical properties of the penalized quantile regression model building on the minimization problem in (\ref{lassoqr}) were widely studied by \cite{BeCh11}. The tuning parameter $\lambda$ acquires a central role, as it controls the intensity of the penalization. Indeed,  
the greater $\lambda$ is, the sparser the solution derived from (\ref{lassoqr}), with an increasing number of coefficients (the portfolio weights) which approach zero.
\cite{BeCh11} introduced a data\textendash driven method with optimal asymptotic properties to determine the optimal value of $\lambda$. This method requires to compute the following quantity:
\begin{equation} \label{Lambda_for_tuning}
\Lambda = T \operatorname*{max}_{\substack{1 \leq j \leq N \\ j\neq s}} \left| \frac{1}{T} \sum_{t=1}^{T} \left[ \frac{x_{j,t}(\tau - \mathbb{I}_{ \{e_t \leq \tau \}}    )}{\widehat{\sigma}_j\sqrt{\tau (1-\tau)}}  \right]   \right|,
\end{equation}
where $e_1, \cdots ,e_T$ are i.i.d. uniform $(0,1)$ random variables.

We then estimate the empirical distribution function of $\Lambda$ by running $B$ iterations and compute the optimal value of $\lambda$ as follows:
\begin{equation}\label{eq:optlambda}
\lambda^\star = c \cdot \Lambda (1-\beta|\textbf{X}), 
\end{equation}
where $\Lambda (1-\beta|\textbf{X})$ is the $(1-\beta)$\textendash th quantile of $\Lambda$ conditional on the covariates $x_{j,t}$, for $1 \leq j \leq N$ and $j \neq s$, whereas $c>1$ is a scalar parameter.

In sum, \cite{BoCaPa} derived the weights of the portfolio with minimum ES at the level $\tau$ from the minimization problem in (\ref{lassoqr}), where the optimal value of $\lambda$ is computed from (\ref{eq:optlambda}). We label this strategy as LBCH, where L and BCH stand, respectively, for the LASSO and for the regularization parameter selection method of \cite{BeCh11}. 

The LASSO is one of the most commonly used penalization method in the statistical literature because it has important properties. Nevertheless, it also suffers from some limitations. For instance, it typically provides biased estimates, overshrinking the retained variables (see e.g., \cite{Fan2001}).
As a result, when implementing the strategy LBCH, the $s$\textendash th asset tends to be overweighted, as highlighted by \cite{BoCaPa}.   
Indeed, the $\ell_1$\textendash norm penalizes the coefficients of the regressors in (\ref{lassoqr}), shrinking the value of $\sum_{\substack{j=1 \\ j\neq s}}^{N} w_j$. This increases the value of $w_s$, being computed in a subsequent step (i.e., after minimizing the loss in (\ref{lassoqr})) as the complement of $\sum_{\substack{j=1 \\ j\neq s}}^{N} w_j$. Moreover, the coefficients derived from (\ref{lassoqr}) are generally downwardly biased in finite samples (see \cite{BeCh11} and \cite{Hautsch14}). 
In this study, we address this issue by using the post\textendash LASSO procedure. This is a well\textendash known method in the financial econometrics literature. For instance, \cite{Hautsch14} used the post\textendash LASSO rule to estimate pessimistic financial networks. However, to the best of our knowledge, this is the first study which uses the post\textendash LASSO estimation to build financial portfolios from penalized (not only quantile) regression models. 

Following \cite{Hautsch14}, we minimize in the first step the loss in (\ref{lassoqr}) and discard the $j$\textendash th regressor $x_{j,t}$ if its shrunken absolute coefficient is sufficiently close to zero; that is, if $|w_{j}| \leq \eta$, where $\eta$ is a given threshold. In contrast, we select the $D$ regressors (with $D \leq N-1$) with a relevant impact\textemdash the ones whose absolute coefficients satisfy the condition $|w_{j}| > \eta$, for $1 \leq j \leq N$ and $j \neq s$. Given the set $\mathcal{D}=\{j_1,\cdots,j_{D}\} \subset \{1,\cdots,N\}$, we define: i) the vector of the selected covariates as $\textbf{x}_{\mathcal{D},t}=\left[x_{j_1,t}\; \cdots\; x_{j_D,t} \right]$; and ii) the vector including their weights as $\textbf{w}_{\mathcal{D}}=\left[w_{j_1}\; \cdots\; w_{j_D} \right]$. 
In the second step, we compute the optimal weights of the selected assets from the following (nonpenalized) minimization problem: 
\begin{equation} \label{unpenqr}
\operatorname*{arg\,min}_{(\textbf{w}_{\mathcal{D}},\;\mu^\star)\in \mathbb{R}^{(D+1)}} \frac{1}{T}\sum_{t=1}^{T} \rho_\tau \left(x_{s,t}- \textbf{x}_{\mathcal{D},t} \cdot  (\textbf{w}_{\mathcal{D}})^\prime -\mu^\star\right),
\end{equation}
whereas the coefficients of the regressors which do not appear in  (\ref{unpenqr}) (i.e., the ones we discard in the first step) are set to zero. 

Therefore, in contrast to \cite{BoCaPa}, we use the LASSO as a selection variable tool only in the first step. Indeed, the portfolio weights are computed in the second step from the nonpenalized minimization problem in (\ref{unpenqr}). The post\textendash LASSO procedure provides improvements under both a statistical and a financial viewpoint. For example, as highlighted by \cite{BeCh11} and \cite{Hautsch14}, the post\textendash LASSO method outperforms both the simple LASSO and the standard quantile regression, which suffer from overidentification problems. Moreover, we compute the portfolio weights without penalizing the coefficients in (\ref{unpenqr}). As a result, we mitigate the complement of the sum $\sum_{d=1}^{D}w_{j_d}$; that is, the weight of the $s$\textendash th asset. We label this strategy as PLBCH, where PL stands for post\textendash LASSO, whereas BCH denotes the method of \cite{BeCh11} that we adopt to select the optimal value of $\lambda$ in the first step.    

The asset allocation strategies described so far rely on the 
regularization parameter selection method of \cite{BeCh11}. Nevertheless, other competing rules are available in the statistical literature. Therefore, it is interesting to compare these different approaches and assess whether and to what extent they affect the portfolio performance. The second selection method we consider in this study is the Bayesian Information Criterion (BIC) introduced by  \cite{Ryung2014}.  \cite{Ryung2014} proposed a BIC  for quantile regression models under the assumption that the error term $\epsilon_t=x_{s,t}- \textbf{x}_{\mathcal{D},t} \cdot  (\widehat{\textbf{w}}_{\mathcal{D}})^\prime -\widehat{\mu}^\star$ has an asymmetric Laplace distribution with the following density function:
\begin{equation}\label{eq:laplacedistr}
f(\epsilon_t; \tau, \sigma) = \frac{\tau(1-\tau)}{\nu} \exp \left\lbrace    -\frac{\rho_\tau(\epsilon_t)}{\nu}\right\rbrace, 
\end{equation} 
where the maximum likelihood estimator of $\nu$ is defined as follows: 
\begin{equation}
\widehat{\nu}=\frac{1}{T}\sum_{t=1}^{T} \rho_\tau \left(x_{s,t}- \textbf{x}_{\mathcal{D},t} \cdot  (\widehat{\textbf{w}}_{\mathcal{D}})^\prime -\widehat{\mu}^\star\right), 
\end{equation}
whereas the coefficients $\widehat{\mu}^\star$ and $\widehat{\textbf{w}}_{\mathcal{D}}$ are obtained by minimizing the loss in (\ref{unpenqr}).

Among the alternative specifications proposed by \cite{Ryung2014}, we adopt the following one:
\begin{equation} \label{BICQR}
BIC_L^H (\mathcal{D}) = \log \left(2 \sum_{t=1}^{T} \rho_\tau \left(x_{s,t}- \textbf{x}_{\mathcal{D},t} \cdot  (\widehat{\textbf{w}}_{\mathcal{D}})^\prime -\widehat{\mu}^\star\right) \right) + |\mathcal{D}| \frac{\log T}{2T} C_T,
\end{equation}
where $|\mathcal{D}|$ is the cardinality of the set $\mathcal{D}=\{j_1,\cdots,j_{D}\}$, whereas $C_T$ is a positive constant. 

We choose $BIC_L^H (\mathcal{D})$ because it provides consistent results in high\textendash dimensional problems \citep{Ryung2014}. Again, we implement the post\textendash LASSO procedure. In particular, we minimize the loss function in (\ref{lassoqr}) by using a large grid of $\lambda$ values. For each value of $\lambda$, we then select the relevant regressors, including them in (\ref{unpenqr}) to obtain the post\textendash LASSO portfolio weights. Among the different solutions derived from (\ref{unpenqr}), we choose the one providing the minimum $BIC_L^H$. We label this strategy as PLBIC because it combines the post\textendash LASSO method with the BIC proposed by \cite{Ryung2014}. 

The third selection method we use in this study is the $K$\textendash fold cross\textendash validation (CV$K$). It is commonly used in applied machine learning, being flexible and easy to implement (see, e.g., \cite{HaTiFr09}). The $K$\textendash fold cross\textendash validation requires the following steps:
\begin{itemize}[noitemsep]
\item[i) ] we randomly divide our dataset into $K$ folds of (approximately) equal size;
\item[ii) ] we use the first fold as the validation set, whereas the remaining ones form the training set;
\item[iii) ] for a given value of $\lambda$ and using the data included in the training set, we minimize the loss in 
(\ref{lassoqr}) and obtain the coefficients 
 $[\widetilde{\mu},\; \widetilde{\textbf{w}}_{(-s)}]$;
\item[iv) ] we assess the fit of the model obtained in iii) on the validation set; that is, we compute the following loss: 
\begin{equation} 
L_1 = \frac{1}{T_1}\sum_{t=1}^{T_1} \rho_\tau \left(x_{s,t}-\sum_{\substack{j=1 \\ j\neq s}}^{N}\widetilde{w}_j x_{j,t}-\widetilde{\mu}\right) \nonumber
\end{equation} 
using the data included in the validation set (the first fold), whose size is equal to $T_1$;
\item[v) ] we repeat the steps ii)\textemdash iv) $K$ times, using, from time to time, the $j$\textendash th fold as the validation set, whereas the others form the training set, for $j=1,...,K$. We then compute the mean of the resulting losses: $\bar{L}=K^{-1}\cdot\sum_{j=1}^{K} L_j$;
\item[vi) ] we repeat the steps ii)\textemdash v) employing a large grid of $\lambda$ values and choose, among them, the value which produces the lowest loss $\bar{L}$. 
\end{itemize}

Other cross\textendash validation schemes are available in the literature. For instance, we cite the single validation set and the leave\textendash one\textendash out cross\textendash validation approaches. However, we choose the $K$\textendash fold cross\textendash validation because it is not computationally expensive and provides accurate estimates \citep{HaTiFr09}. 
As said above, the LASSO suffers from relevant estimation problems \citep{Fan2001}.         
This has also motivated the development of other penalty functions. Two relevant examples are the Smoothly Clipped Absolute Deviation (SCAD) and the Minimax Concave Penalty (MCP), introduced by \cite{Fan2001} and \cite{Zhang2010}, respectively. In contrast to the LASSO, the SCAD and the MCP build on nonconvex penalty functions and possess the oracle property as defined by \cite{Fan2001}. 
%
%
%
In particular, when using the SCAD, we solve the following minimization problem: 
\begin{equation} \label{scadqr}
\operatorname*{arg\,min}_{(\textbf{w}_{(-s)},\;\mu)\in \mathbb{R}^N} \frac{1}{T}\sum_{t=1}^{T} \rho_\tau \left(x_{s,t}-\sum_{\substack{j=1 \\ j\neq s}}^{N}w_j x_{j,t}-\mu\right)+ 
\end{equation}
$$
+ \sum_{\substack{j=1 \\ j\neq s}}^{N} \lambda |w_j|\mathbb{I}_{\{0\leq |w_j| < \lambda \}} + \frac{a \lambda  |w_j|-\left(w_j^2+\lambda^2\right)/2}{a-1} \mathbb{I}_{\{\lambda \leq |w_j| \leq a\lambda \}} + \frac{\left(a+1\right)\lambda^2}{2}\mathbb{I}_{\{|w_j| > a\lambda \}},
$$
where $a>2$ and $\lambda>0$.

In contrast, the minimization problem we solve when using the MCP is defined as follows: 
\begin{equation} \label{mcpqr}
\operatorname*{arg\,min}_{(\textbf{w}_{(-s)},\;\mu)\in \mathbb{R}^N} \frac{1}{T}\sum_{t=1}^{T} \rho_\tau \left(x_{s,t}-\sum_{\substack{j=1 \\ j\neq s}}^{N}w_j x_{j,t}-\mu\right)+ 
\end{equation}
$$
+ \sum_{\substack{j=1 \\ j\neq s}}^{N} \lambda \left(|w_j|-\frac{w_j^2}{2a\lambda}\right) \mathbb{I}_{\{0\leq |w_j| \leq a\lambda \}} + \frac{a\lambda^2}{2}\mathbb{I}_{\{|w_j| > a\lambda \}},
$$
where $a>1$ and $\lambda>0$.

We employ the $K$\textendash fold cross\textendash validation method on the three penalty functions described above; that is, LASSO, SCAD and MCP. In doing so, we can evaluate the impact of the different penalty specifications for a given selection method (CV$K$), with or without the post\textendash penalization procedure. In the former case, we select the relevant regressors from (\ref{lassoqr}), (\ref{scadqr}) and (\ref{mcpqr}), respectively, using the selection method CV$K$. We then compute the portfolio weights from (\ref{unpenqr}). The corresponding strategies are denoted as PLCV$K$ (post\textendash LASSO with CV$K$), PSCV$K$ (post\textendash SCAD with CV$K$) and PMCV$K$ (post\textendash MCP with CV$K$). In contrast, the portfolio weights are directly computed from (\ref{lassoqr}), (\ref{scadqr}) and (\ref{mcpqr}) when we do not use the post\textendash penalization method. The resulting investment rules are then denoted as LCV$K$, SCV$K$ and PMCV$K$, respectively.

\section{Empirical set\textendash up}\label{sec:empsetup}
We implement the portfolio strategies described in Section \ref{sec:methods} using two datasets: the 49 Industry Portfolios (49P) and the 100 Portfolios Formed on Size and Book\textendash to\textendash Market (100P).\footnote{We recover the data from the website http://mba.tuck.dartmouth.edu/pages/faculty/ken.french/ using the library provided by Kenneth R. French. Starting from the original daily data, we compute the weekly returns for both datasets.} We use a weekly horizon as it better reflects the rebalancing activity of fund managers, while a daily frequency is typically a too short time interval. Both datasets include $Q=1,006$ weeks in the interval 14/01/2000\textemdash 26/04/2019. The number of assets ($N$) is equal to 49 when using the 49P dataset, whereas $N=100$ in the 100P dataset. We focus on the out\textendash of\textendash sample performance rather than on the in\textendash sample one because only the former reproduces the real activity of investors who rebalance weekly their portfolios.\footnote{The in\textendash sample results are available upon request.} The evaluation of the out\textendash of\textendash sample performance builds on a rolling window procedure, that we describe below. 

For both datasets, we iteratively divide the overall time series with a dimension $Q \times N$ into $Q-T$ equally\textendash sized subsamples, each of which has a dimension $T \times N$, where $T<Q$. As a result, the first subsample includes the returns from the first to the $T$\textendash th week. The second subsample is obtained by removing the oldest observations and including the ones of the $(T+1)$\textendash th week. This procedure continues until the $(Q-1)$\textendash th week is reached, with the last subsample which spans the interval $[Q-T, Q-1]$. We use two different values of $T$ in the empirical analysis; that is, $T=\{100,200\}$. This allows us to check whether the results change according to the portfolio dimensionality ($N$) as well as to the sample size ($T$). The case of $T=100$ and $N=100$ acquires a relevant interest in this study, as it allows us to test the effectiveness of the regularization techniques when the ratio $T/N$ takes relatively low values. For each subsample and for a given investment rule, we estimate the portfolio weights denoted as $\widehat{\textbf{w}}_t$, for $t=T,...,Q-1$. We then obtain the out\textendash of\textendash sample portfolio returns as $r_{p,t+1}=\textbf{r}_{t+1} \widehat{\textbf{w}}_t^\prime$, for $t=T,...,Q-1$. As a result, we obtain a $(Q-T)$ vector of out\textendash of\textendash sample portfolio returns for each investment strategy, from which we compute the performance measures described below. Following \cite{Ch08}, we first compute the expected shortfall at the level $\tau$ as follows:
\begin{equation}\label{eq:es}
ES = - \frac{\sum_{t=T}^{Q-1} r_{p,t+1} \mathbb{I}_{\{r_{p,t+1} < VaR  \}}   }{ \sum_{t=T}^{Q-1}\mathbb{I}_{\{r_{p,t+1} < VaR  \}} },
\end{equation}
where $VaR$ is the value\textendash at\textendash risk of the portfolio return; that is, the $\tau$\textendash th quantile of 
$r_{p,t+1}$, for $t=T,...,Q-1$.

The second measure we compute is the sample standard deviation of $r_{p,t+1}$, for $t=T,...,Q-1$, denoted as $SD$. Assuming that the risk\textendash free rate is equal to zero, we then calculate the Sharpe ratio (SR) as follows:
\begin{equation}\label{eq:sharperatio}
SR = \frac{\bar{r}_p}{SD}, 
\end{equation}
where $\bar{r}_p$ is the sample mean of $r_{p,t+1}$, for $t=T,...,Q-1$.

In the empirical analysis, we also test whether the variances and the Sharpe ratios of the competing investment strategies are statistically different using the test proposed by \cite{LeWo08}. The three statistics described above are computed on the portfolio returns and convey information about the extreme risk (i.e., the risk of incurring extreme losses when tail events occur), the volatility and the risk\textendash adjusted return. In contrast, the statistics described below are computed on the portfolio weights.  We evaluate the stability of the estimates (i.e., the portfolio weights) as well as the impact of the trading fees on the rebalancing activity using the turnover, defined as follows:
\begin{equation}\label{eq:turnover}
TO= \frac{1}{Q-T-1} \sum_{t=T+1}^{Q-1}  \left[ \sum_{j=1}^{N} |\widehat{w}_{j,t}-w^\star_{j,t}| \right],
\end{equation}
where $\widehat{w}_{j,t}$ and $w^\star_{j,t}$ are the weights of the $j$\textendash th asset at time $t$, immediately after and immediately before rebalancing the portfolio, respectively, for $j=1,...,N$. 

We also compute the average number of active ($AP$) and short ($SP$) positions as follows: 
\begin{equation} \label{eq:actpos}
AP = \frac{1}{Q-T}\sum_{t=T}^{Q-1}\left[ \sum_{j=1}^{N} \mathbb{I}_{\{|\widehat{w}_{j,t}| > \eta\} } \right],
\end{equation}
\begin{equation} \label{eq:shortpos}
SP = \frac{1}{Q-T}\sum_{t=T}^{Q-1}\left[ \sum_{j=1}^{N} \mathbb{I}_{\{\widehat{w}_{j,t} < -\eta\} } \right],
\end{equation}
where $\eta$ is the same threshold value that we impose when selecting the relevant covariates to employ in the post\textendash penalization procedure (see Section \ref{sec:methods}); we then focus on either active or short positions which are significantly different from zero; in our empirical study, we set $\eta=0.00001$. 

We report below a set of details about the implementation of the investment strategies listed in Table \ref{tab:liststrategies}.
We set $\tau=0.05$ when estimating the regression models described in Section \ref{sec:methods} as well as when computing ES in (\ref{eq:es}). In doing so, we focus on the left\textendash tail relationships among the assets we focus on. 
As we already said in Section \ref{sec:methods}, selecting the value of $s\in \{1,...,N\}$ to define the response variable $x_{s,t}$ of the regression models defined in Section \ref{sec:methods} is an important step. Following \cite{BoCaPa}, we define $x_{s,t}$ as the return yielded by the $j$\textendash th asset ($1 \leq j \leq N$) which records the lowest in\textendash sample expected shortfall for each subsample in the rolling window scheme. In doing so, we emphasize the central role of the expected shortfall in this study\textemdash it is the target measure we minimize when estimating a quantile regression model\textemdash and always select the asset with the lowest extreme risk, providing benefits for the overall portfolio. Following the recommendation  of \cite{BeCh11}, we set $B=1,000$, $c=2$ and $1-\beta=0.9$ when implementing their regularization parameter selection method. When using the BIC of \cite{Ryung2014}, we set $C_T=\log T$ because this choice provides good results in a wide range of settings \citep{Ryung2014}. In contrast, we implement the $K$\textendash fold cross\textendash validation by setting $K=5$, which  represents a standard choice in the literature \citep{HaTiFr09}.
We employ both BIC and CV5 using a large grid of 100 values of $\lambda$, from which we select the optimal tuning parameter. 
We estimate the quantile regression models defined in Section \ref{sec:methods} using the R package `quantreg' when using the LASSO. In contrast, we use the R package `rqPen' when using the penalties SCAD and MCP. We check that the LASSO penalty, combined with the selection method BCH, significantly reduces the computational burden.

For instance, when implementing the rule LBCH, the mean runtimes we record on an Intel\textregistered Core i7-4710HQ@2.50GHz (64-bit operating system) computer for a single $T \times N$ subsample are equal to: 0.031 (49P dataset with $T=100$), 0.047 (49P dataset with $T=200$), 0.047 (100P dataset with $T=100$) and 0.062 (100P dataset with $T=200$) seconds. These values increase to 0.469, 0.621, 1.718 and 2.167 seconds, respectively, when adopting the strategy LCV5 (we record similar runtimes applying the selection method BIC). They significantly increase when using either the SCAD or the MCP penalty functions. For example, we record the following mean runtimes for the rule SCV5: 2.546 (49P dataset with $T=100$), 3.343 (49P dataset with $T=200$), 6.953 (100P dataset with $T=100$) and 32.283 (100P dataset with $T=200$) seconds (we observe similar values for the strategy MCV5). Finally, we check that the runtimes do not significantly increase when adopting a post\textendash penalization procedure.

\begin{table}[htbp]
\setlength{\tabcolsep}{26pt}
\small
\begin{center}
  \caption{Investment strategies}
   \begin{tabular}{cccc}
    \toprule
    LABEL & post\textendash penalization & PENALTY & SELECTION METHOD\\
    \midrule
    LBCH  & N     & LASSO &  BCH\\
    LCV5  & N & LASSO & CV5 \\
    PLBCH & Y     & LASSO &  BCH\\
    PLCV5 & Y     & LASSO &  CV5\\
    PLBIC & Y     & LASSO &  BIC\\
    SCV5 & N     & SCAD &  CV5 \\
    PSCV5 & Y     & SCAD &  CV5\\
    MCV5 & N     & MCP &  CV5 \\
    PMCV5 & Y     & MCP &  CV5\\
    EW &  \textemdash    & \textemdash & \textemdash\\  
    \bottomrule
    \end{tabular}%

    \label{tab:liststrategies}
    \end{center}
  \bigskip
The table reports the description of the asset allocation strategies we implement in the empirical analysis. The first column reports the label of each strategy. The second column indicates whether the post\textendash penalization procedure has been used (Y) or not (N) to compute the final portfolio weights. The third column reports the specification of the penalty function. The fourth column reports the regularization parameter selection method. We use the following selection methods: i) BCH; that is, the method of \cite{BeCh11}; ii) CV5; that is, the 5\textendash fold cross\textendash validation; and iii) BIC; that is, the Bayesian Information Criterion of \cite{Ryung2014}.
EW is the equally weighted portfolio.
\end{table}%

\section{Empirical findings}\label{sec:empirical_results}
We report the out\textendash of\textendash sample results in Table \ref{tab:oosresults}. We first analyze the performance of the competing asset allocation strategies in terms of expected shortfall (ES), as it is the target measure we aim to minimize when building financial portfolios from quantile regression models. PLBCH records the best performance, generating the lowest ES in all but one case, whereas only PLCV5 outperforms 
PLBCH when using the 49P dataset with $T=200$. We now analyze in more detail the drivers of this performance, which depends on more than one factor, such as the specification of the penalty function, the regularization parameter selection method and the use of the post\textendash penalization procedure. We first compare LCV5, SCV5 and MCV5 to evaluate the impact of the different penalty functions for a given selection criterion (CV5), excluding the effects of the post\textendash penalization procedure. MCV5 generates the lowest ES in all but one case. However, LCV5, SCV5 and MCV5 produce similar values of ES. In contrast, we record significant improvements when applying the post\textendash penalization procedure. Indeed, the values of ES significantly decrease when moving from LBCH, LCV5, SCV5 and MCV5 to PLBCH, PLCV5, PSCV5 and PMCV5, respectively. We now compare the contribution of the different selection criteria, focusing on PLBCH, PLCV5 and PLBIC. The method of \cite{BeCh11}\textemdash BCH\textemdash outperforms the other criteria in all but one case (with the only exception in Panel (d) of Table  \ref{tab:oosresults}, where PLBCH achieves the second best performance). Interestingly, the good performance of BCH is less evident when the post\textendash LASSO method is not used, as LCV5 outperforms LBCH in all but one case. We highlight 
the gap between EW and the other portfolios (which are retrieved from penalized quantile regression models), emphasizing the utility provided by the latter when the extreme risk is the primary focus.

\begin{table}[htbp]
\setlength{\tabcolsep}{4,5pt}
\small
\begin{center}
  \caption{Out\textendash of\textendash sample statistics}
    \begin{tabular}{c|cccccc|cccccc}
    \toprule
    STRATEGY & ES    & SD    & SR    & TO    & AP    & SP    & ES    & SD    & SR    & TO    & AP    & SP \\
    \midrule
    \multicolumn{1}{r}{} & \multicolumn{6}{|c|}{(a) 49P DATASET, $T=100$}       & \multicolumn{6}{c}{(b) 49P DATASET, $T=200$} \\
    \midrule
    LBCH  & 4.378 & 1.877 & 9.334 & 0.143 & 3.861 & 1.103 & 4.288 & 1.881 & 11.917 & 0.108 & 5.483 & 1.679 \\
    LCV5  & 4.492 & 1.900 & 8.767 & 0.430 & 5.075 & 1.840 & 4.236 & 1.854 & 9.314 & 0.585 & 10.295 & 3.603 \\
    PLBCH & 3.150 & 1.291 & 12.060 & 0.283 & 2.827 & 0.929 & 3.894 & 1.723 & 13.901 & 0.240 & 5.238 & 1.716 \\
    PLCV5 & 4.000 & 1.636 & 7.196 & 0.840 & 4.634 & 1.708 & 3.680 & 1.671 & 11.259 & 1.068 & 10.239 & 3.743 \\
    PLBIC & 4.010 & 1.634 & 4.859 & 0.207 & 2.040 & 0.762 & 4.017 & 1.613 & 8.989 & 0.262 & 2.895 & 1.029 \\
    SCV5  & 4.450 & 1.883 & 7.994 & 0.652 & 4.230 & 1.477 & 4.473 & 1.937 & 9.753 & 1.028 & 7.091 & 2.645 \\
    PSCV5 & 3.825 & 1.514 & 7.835 & 0.825 & 3.619 & 1.323 & 3.950 & 1.716 & 11.057 & 1.102 & 6.883 & 2.603 \\
    MCV5  & 4.408 & 1.875 & 7.632 & 0.641 & 4.236 & 1.481 & 4.450 & 1.945 & 9.686 & 1.003 & 7.048 & 2.632 \\
    PMCV5 & 3.784 & 1.523 & 7.154 & 0.814 & 3.632 & 1.337 & 3.912 & 1.722 & 11.168 & 1.080 & 6.831 & 2.587 \\
    EW    & 6.376 & 2.546 & 6.137 & 0.015 & 49.000 & 0.000 & 6.475 & 2.578 & 6.273 & 0.015 & 49.000 & 0.000 \\
    \midrule
    \multicolumn{1}{c}{} & \multicolumn{6}{|c|}{(c) 100P DATASET, $T=100$}      & \multicolumn{6}{c}{(d) 100P DATASET, $T=200$} \\
    \midrule
    LBCH  & 4.870 & 2.128 & 13.251 & 0.228 & 4.188 & 1.983 & 4.819 & 2.010 & 8.727 & 0.189 & 5.797 & 3.022 \\
    LCV5  & 4.724 & 2.101 & 13.619 & 0.753 & 6.205 & 2.955 & 4.042 & 1.782 & 13.014 & 1.163 & 15.167 & 7.743 \\
    PLBCH & 3.642 & 1.514 & 12.504 & 0.563 & 3.210 & 1.600 & 3.744 & 1.609 & 10.061 & 0.472 & 5.220 & 2.635 \\
    PLCV5 & 4.177 & 1.824 & 11.986 & 1.423 & 5.820 & 2.843 & 3.889 & 1.702 & 14.881 & 2.084 & 15.159 & 7.728 \\
    PLBIC & 3.803 & 1.656 & 10.502 & 0.204 & 1.359 & 0.627 & 4.574 & 1.846 & 7.121 & 0.437 & 3.618 & 1.764 \\
    SCV5  & 4.767 & 2.119 & 14.619 & 1.424 & 7.882 & 3.926 & 4.022 & 1.757 & 12.015 & 2.230 & 18.543 & 9.603 \\
    PSCV5 & 4.387 & 1.864 & 10.399 & 2.269 & 7.523 & 3.712 & 3.926 & 1.757 & 13.226 & 3.561 & 18.530 & 9.561 \\
    MCV5  & 4.723 & 2.135 & 14.144 & 1.476 & 8.215 & 4.147 & 3.915 & 1.729 & 13.076 & 2.161 & 18.737 & 9.649 \\
    PMCV5 & 4.198 & 1.855 & 12.947 & 2.447 & 7.861 & 3.911 & 3.846 & 1.735 & 13.422 & 3.593 & 18.721 & 9.643 \\
    EW    & 6.872 & 2.785 & 5.864 & 0.010 & 100.000 & 0.000 & 7.011 & 2.810 & 5.854 & 0.010 & 100.000 & 0.000 \\
    \bottomrule
    \end{tabular}%
    \label{tab:oosresults}
    \end{center}
  \bigskip
The table reports the performance measures computed for the portfolio strategies listed in the first column. From left to right, the table reports the following statistics: expected shortfall at the 5\% level (ES, \%), standard deviation (SD, \%), Sharpe ratio (SR, \%), turnover (TO), mean of active (AP) and short (SP) positions. These statistics are computed by using two different datasets: the 49 Industry Portfolios (49P) and the 100 Portfolios Formed on Size and Book\textendash to\textendash Market (100P), implementing the rolling window procedure with two different sizes of the estimation window ($T=\{100,200\}$). 
\end{table}%

We obtain similar results when analyzing the standard deviation (SD) of the out\textendash of\textendash sample portfolio returns. Indeed, PLBCH records the best performance, providing the lowest standard deviation in all but one case (with the only exception in Panel (b) of Table  \ref{tab:oosresults}, where PLBCH achieves the second best performance). Moreover, the standard deviations produced by PLBCH and the ones obtained from the other strategies are statistically different at the 5\% level in most of the cases (see the results of the \cite{LeWo08}'s test in Panel (a) of Tables \ref{tab:TESTLW49P} and \ref{tab:TESTLW100P}).

\begin{table}[htbp]
\setlength{\tabcolsep}{6pt}
\small
\begin{center}
  \caption{\cite{LeWo08}'s test using the 49P dataset}
    \begin{tabular}{c|cccccccccc}
    \toprule
    \multicolumn{11}{c}{(a) VARIANCE} \\
    \midrule
          & LBCH  & LCV5  & PLBCH & PLCV5 & PLBIC & SCV5  & PSCV5 & MCV5  & PMCV5 & EW \\
    \midrule
    LBCH  & -     & 27.237 & 0.000 & 0.112 & 0.208 & 86.812 & 0.000 & 93.938 & 0.000 & 0.000 \\
    LCV5  & 38.033 & -     & 0.000 & 0.009 & 0.046 & 57.838 & 0.000 & 40.993 & 0.000 & 0.000 \\
    PLBCH & 0.002 & 0.076 & -     & 0.214 & 0.507 & 0.000 & 5.032 & 0.000 & 4.129 & 0.000 \\
    PLCV5 & 5.862 & 10.585 & 63.518 & -     & 97.436 & 0.038 & 5.635 & 0.057 & 8.388 & 0.000 \\
    PLBIC & 2.871 & 6.183 & 38.741 & 40.671 & -     & 0.193 & 9.608 & 0.274 & 12.390 & 0.000 \\
    SCV5  & 27.060 & 9.823 & 0.021 & 2.103 & 0.752 & -     & 0.000 & 18.039 & 0.000 & 0.000 \\
    PSCV5 & 18.752 & 28.840 & 95.619 & 46.305 & 8.163 & 4.852 & -     & 0.000 & 37.538 & 0.000 \\
    MCV5  & 19.891 & 7.469 & 0.023 & 1.703 & 0.553 & 46.858 & 4.088 & -     & 0.000 & 0.000 \\
    PMCV5 & 20.330 & 31.028 & 99.336 & 41.520 & 6.854 & 5.674 & 65.528 & 4.526 & -     & 0.000 \\
    EW    & 0.000 & 0.000 & 0.000 & 0.000 & 0.000 & 0.000 & 0.000 & 0.000 & 0.000 & - \\
    \midrule
    \multicolumn{11}{c}{(b) SHARPE RATIO} \\
    \midrule
          & LBCH  & LCV5  & PLBCH & PLCV5 & PLBIC & SCV5  & PSCV5 & MCV5  & PMCV5 & EW \\
    \midrule
    LBCH  & -     & 54.392 & 37.209 & 37.067 & 6.273 & 39.553 & 57.093 & 26.071 & 39.654 & 24.830 \\
    LCV5  & 3.303 & -     & 29.432 & 46.415 & 9.413 & 57.439 & 70.456 & 39.423 & 50.056 & 38.210 \\
    PLBCH & 22.082 & 1.189 & -     & 16.275 & 2.820 & 18.311 & 22.110 & 15.083 & 15.095 & 11.703 \\
    PLCV5 & 79.544 & 37.491 & 34.951 & -     & 37.488 & 73.018 & 79.610 & 84.895 & 98.607 & 76.886 \\
    PLBIC & 33.568 & 91.580 & 13.683 & 44.384 & -     & 20.002 & 30.551 & 25.783 & 42.330 & 71.317 \\
    SCV5  & 25.451 & 78.198 & 6.899 & 57.917 & 80.700 & -     & 94.166 & 30.145 & 69.935 & 56.759 \\
    PSCV5 & 75.899 & 52.152 & 36.704 & 93.507 & 44.833 & 53.039 & -     & 92.658 & 37.237 & 64.397 \\
    MCV5  & 25.393 & 81.935 & 7.012 & 56.508 & 82.548 & 90.673 & 52.226 & -     & 82.292 & 64.281 \\
    PMCV5 & 79.331 & 50.018 & 39.270 & 97.150 & 42.811 & 51.611 & 88.907 & 47.371 & -     & 78.113 \\
    EW    & 6.484 & 36.351 & 2.564 & 19.716 & 53.075 & 33.597 & 23.874 & 35.038 & 23.217 & - \\
    \bottomrule
    \end{tabular}%
    \label{tab:TESTLW49P}
    \end{center}
  \bigskip
For each pair of the portfolio strategies ordered by row and column, the table provides the p\textendash values (\%) of the \cite{LeWo08}'s test that we implement to check whether the differences computed on both the variances (Panel (a)) and Sharpe ratios (Panel (b)) are statistically significant. The results are obtained by using the 49 Industry Portfolios (49P) dataset. We report the p\textendash values obtained by setting $T=100$ ($T=200$) in the rolling window procedure above (below) the main diagonals in Panels (a)\textemdash (b). 
\end{table}%

Again, we do not observe relevant differences when comparing the different penalty functions in LCV5, SCV5 and MCV5 and none of these three rules dominates the others. In fact, LCV5 records the best performance when using the 49P dataset with $T=200$ and the 100P dataset with $T=100$, whereas MCV5 provides the lowest SD in the remaining cases; SCV5 always achieves the second best performance. The post\textendash penalization procedure reduces the values of SD in all but two cases (the volatility does not decrease from SCV5 and MCV5 to PSCV5 and PMCV5, respectively, in Panel (d) of Table \ref{tab:oosresults}). As with the previous analysis, the selection method of \cite{BeCh11} provides the best results when using the post\textendash LASSO procedure (PLBCH outperforms PLCV5 and PLBIC in all but one case), whereas this evidence is less clear without the post\textendash LASSO (for instance, when comparing LBCH with LCV5). The gap between EW and the other rules is relevant also in terms of SD, as the former provides portfolios which are much more volatile. In addition, the differences between EW and the other strategies are always highly statistically significant (see Panel (a) of Tables \ref{tab:TESTLW49P} and \ref{tab:TESTLW100P}).

\begin{table}[htbp]
\setlength{\tabcolsep}{6pt}
\small
\begin{center}
  \caption{\cite{LeWo08}'s test using the 100P dataset}
    \begin{tabular}{c|cccccccccc}
    \toprule
    \multicolumn{11}{c}{(a) VARIANCE} \\
    \midrule
          & LBCH  & LCV5  & PLBCH & PLCV5 & PLBIC & SCV5  & PSCV5 & MCV5  & PMCV5 & EW \\
    \midrule
    LBCH  & -     & 48.651 & 0.000 & 0.000 & 0.000 & 82.664 & 0.243 & 88.578 & 0.002 & 0.000 \\
    LCV5  & 0.049 & -     & 0.000 & 0.000 & 0.000 & 52.176 & 0.365 & 35.152 & 0.000 & 0.000 \\
    PLBCH & 0.002 & 0.394 & -     & 0.251 & 21.523 & 0.000 & 0.253 & 0.000 & 0.123 & 0.000 \\
    PLCV5 & 0.002 & 13.234 & 13.941 & -     & 1.658 & 0.000 & 59.300 & 0.000 & 50.844 & 0.000 \\
    PLBIC & 3.079 & 39.353 & 0.719 & 1.541 & -     & 0.000 & 2.106 & 0.000 & 0.288 & 0.000 \\
    SCV5  & 0.007 & 61.021 & 2.299 & 12.310 & 8.114 & -     & 0.133 & 68.513 & 0.000 & 0.000 \\
    PSCV5 & 0.129 & 69.308 & 3.472 & 9.065 & 15.177 & 99.247 & -     & 0.107 & 88.837 & 0.000 \\
    MCV5  & 0.002 & 34.157 & 8.035 & 41.116 & 3.139 & 22.336 & 44.671 & -     & 0.000 & 0.000 \\
    PMCV5 & 0.058 & 44.778 & 7.351 & 29.367 & 7.493 & 56.928 & 42.841 & 86.000 & -     & 0.000 \\
    EW    & 0.000 & 0.000 & 0.000 & 0.000 & 0.000 & 0.000 & 0.000 & 0.000 & 0.000 & - \\
    \midrule
    \multicolumn{11}{c}{(b) SHARPE RATIO} \\
    \midrule
          & LBCH  & LCV5  & PLBCH & PLCV5 & PLBIC & SCV5  & PSCV5 & MCV5  & PMCV5 & EW \\
    \midrule
    LBCH  & -     & 74.836 & 79.468 & 57.863 & 31.212 & 32.747 & 29.670 & 55.666 & 90.404 & 0.717 \\
    LCV5  & 2.305 & -     & 68.865 & 42.232 & 23.247 & 37.651 & 22.894 & 68.547 & 78.145 & 0.591 \\
    PLBCH & 61.385 & 27.823 & -     & 86.902 & 56.815 & 48.790 & 55.017 & 58.748 & 89.619 & 7.432 \\
    PLCV5 & 0.999 & 18.238 & 9.413 & -     & 55.645 & 21.871 & 52.689 & 34.496 & 66.666 & 6.894 \\
    PLBIC & 51.585 & 1.475 & 33.225 & 0.421 & -     & 12.042 & 97.196 & 18.842 & 36.616 & 17.628 \\
    SCV5  & 10.825 & 48.498 & 48.246 & 8.573 & 4.313 & -     & 8.350 & 64.948 & 45.467 & 0.246 \\
    PSCV5 & 7.333 & 91.265 & 28.084 & 33.589 & 3.116 & 40.361 & -     & 15.190 & 18.056 & 17.532 \\
    MCV5  & 4.281 & 96.675 & 29.489 & 26.600 & 1.762 & 33.727 & 92.909 & -     & 58.152 & 0.491 \\
    PMCV5 & 7.775 & 84.275 & 28.293 & 44.147 & 3.126 & 44.304 & 89.757 & 81.859 & -     & 3.002 \\
    EW    & 30.201 & 2.247 & 22.521 & 0.430 & 70.960 & 5.254 & 2.835 & 2.937 & 2.833 & - \\
    \bottomrule
    \end{tabular}%
    \label{tab:TESTLW100P}
    \end{center}
  \bigskip
For each pair of the portfolio strategies ordered by row and column, the table provides the p\textendash values (\%) of the \cite{LeWo08}'s test that we implement to check whether the differences computed on both the variances (Panel (a)) and Sharpe ratios (Panel (b)) are statistically significant. The results are obtained by using the 100 Portfolios Formed on Size and Book\textendash to\textendash Market (100P) dataset. We report the p\textendash values obtained by setting $T=100$ ($T=200$) in the rolling window procedure above (below) the main diagonals in Panels (a)\textemdash (b).  
\end{table}%

We now analyze the Sharpe ratio (SR) of the out\textendash of\textendash sample portfolio returns. We can see from Table \ref{tab:oosresults} that PLBCH provides the best results when using the 49P dataset, for both sample sizes ($T=\{100,200\}$). Nevertheless, the differences with the other strategies are rarely statistically significant (see Panel (b) of Table \ref{tab:TESTLW49P}). In contrast, SCV5 outperforms the other rules when using the 100P dataset with $T=100$, whereas MCV5 achieves the second best performance. This highlights the good performance of SCV5 and MCV5 in terms of risk\textendash adjusted return for low values of the ratio $T/N$, despite SCV5 and MCV5 suffer from relatively high standard deviations. PLCV5 yields the greatest SR when using the 100P dataset with $T=200$, whereas PMCV5 and PSCV5 record the second and the third best results, outperforming MCV5 and SCV5, respectively. However, the differences with the other strategies are rarely statistically significant (see Panel (b) of Table \ref{tab:TESTLW100P}). When comparing the different penalty functions, we check that LCV5, SCV5 and MCV5 generate similar values of SR, consistent with the previous analyses. The effects of the post\textendash penalization procedure are less clear with respect to the previous analyses. In fact, PLBCH outperforms LBCH in all but one case, whereas PLCV5, PSCV5 and PMCV5 outperform LCV5, SCV5 and MCV5, respectively, only for $T=200$. Again, the selection method of \cite{BeCh11} works better when using the post\textendash LASSO procedure. Indeed, PLBCH outperforms PLCV5 and PLBIC in all but one case. In contrast, LCV5 dominates LBCH when using the 100P dataset. The portfolios derived from the penalized quantile regression models outperform EW in all but one case and the resulting differences are often statistically significant at the 5\% level, mainly when using the 100P dataset (see Panel (b) of Tables \ref{tab:TESTLW49P} and \ref{tab:TESTLW100P}).

We now analyze the statistics computed on the portfolio weights, starting from the turnover (TO). EW outperforms the other strategies with values of TO that approach zero (see Table \ref{tab:oosresults}). This is due to the fact that the optimal weights of EW that we set when rebalancing the portfolio are constant at the level $1/N$. When restricting the attention on the other strategies derived from penalized quantile regression models, we check that the rules in which we combine the LASSO penalty with either the BCH \citep{BeCh11} or the BIC \citep{Ryung2014} selection criteria (i.e., the strategies LBCH, PLBCH and PLBIC) provide values of TO which are significantly lower than the ones obtained from the other portfolios. Therefore, LBCH, PLBCH and PLBIC provide advantages in terms of 
a greater stability in the resulting estimates and a lower impact of transaction costs.

In contrast to EW, in which all positions are active and long by definition, the portfolios derived from penalized quantile regression models are sparse (see Table \ref{tab:oosresults}). Indeed, they yield, on average, 3.795 (6.889) active positions (AP) when using the 49P dataset with $T=100$ ($T=200$). These values increase to 5.807 (13.277) when using the 100P dataset with $T=100$ ($T=200$). Therefore, the lower the ratio $T/N$ is, the greater the impact of the penalty in the underlying quantile regression models, resulting in a lower number of active positions. The strategies in which we use either the BCH \citep{BeCh11} or the BIC \citep{Ryung2014} criteria (i.e., LBCH, PLBCH and PLBIC) provide, on average, a lower number of active positions, consistent with the values of TO. Moreover, they also yield, on average, a lower number of short positions (SP). As a result, LBCH, PLBCH and PLBIC would be useful in situations where short\textendash selling restrictions 
are imposed.

\begin{figure}[htb!]
\hbox{\hspace{-2cm}\includegraphics[scale=0.44]{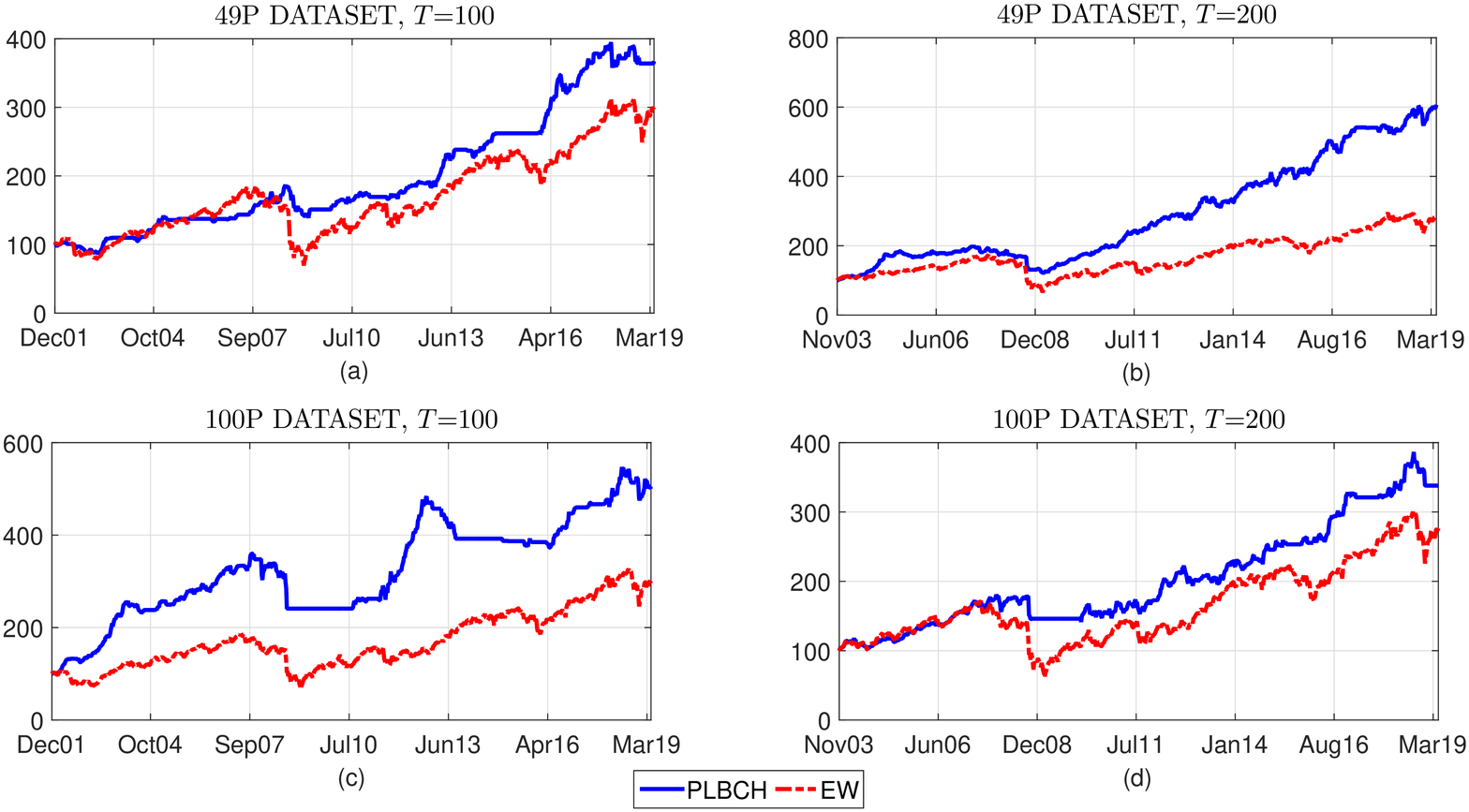}}
\caption{\footnotesize{The figure displays the wealth provided by PLBCH and EW over time. Subfigures (a)\textemdash (b) display the results we obtain using the 49 Industry Portfolios (49P) dataset, implementing the rolling window procedure with, respectively, $T=100$ and $T=200$  observations. Subfigures (c)\textemdash (d) display the results we obtain using the 100 Portfolios Formed on Size and Book\textendash to\textendash Market (100P) dataset, implementing the rolling window procedure with, respectively, $T=100$ and $T=200$  observations.}}
\label{fig:trend_wealth}
\end{figure}

In sum, a joint analysis of the six performance indicators described above, which convey information about the portfolio performance from different viewpoints, such as risk, profitability and turnover, highlights the significant improvements provided by the post\textendash penalization procedure. In addition, the results improve further when combining the post\textendash LASSO with the selection method of \cite{BeCh11} within the same strategy; that is, PLBCH. This evidence is stronger when focusing on the risk dimension, in terms of both standard deviation (the results acquire a greater statistical significance with the \cite{LeWo08}'s test) and expected shortfall. In particular, the latter has a relevant role in this study, as it represents the measure we minimize when estimating a quantile regression model. Surprisingly, PLBCH outperforms EW, a benchmark known to be difficult to beat out\textendash of\textendash sample (see, e.g., \cite{DeGaUp09}, \cite{DuLe09} and \cite{TuZh11}), also in terms of risk\textendash adjusted return (quantified by the Sharpe ratio in Table \ref{tab:oosresults}) and profitability. This evidence is clear when comparing the wealth produced by PLBCH and EW, respectively. Indeed, PLBCH dominates EW, yielding a greater wealth in most of the weeks included in our dataset (see Figure \ref{fig:trend_wealth}).

Finally, we check that the post\textendash penalization procedure turns out to be an effective tool to deal with the portfolio concentration issues highlighted by \cite{BoCaPa}. Indeed, it significantly mitigates the weight of the $s$\textendash th asset; that is, the one whose return represents the response variable in the estimated quantile regression model. For instance, when using the 49P dataset with $T=100$, such a weight falls from an average value of 0.9310 with LBCH to an average value of 0.3621 with PLBCH.\footnote{The results obtained with the 100P dataset and for both sizes of the estimation window ($T$) are similar and available upon request.}


\section{Concluding remarks}\label{sec:conclusions}
We show that the performance of financial portfolios derived from penalized quantile regression models crucially depends on the specification of the penalty function, on the regularization parameter method selection and on the use of the post\textendash penalization procedure. Although the properties of these methods have been widely studied in the statistical literature, they have never been tested from a financial viewpoint when building investment strategies using quantile regression models.    

A joint analysis of six performance indicators providing information about risk, profitability and turnover highlights the significant improvements we can achieve going beyond the simple LASSO method. These improvements are clear when focusing on the risk dimension, in terms of a lower volatility and extreme risk. This is a relevant result in this study, as the expected shortfall represents the target measure we aim to minimize when estimating a quantile regression model. Surprisingly, we 
also record gains in terms of risk\textendash adjusted return and profitability, outperforming the equally weighted portfolio, which is a benchmark known to be difficult to beat out\textendash of\textendash sample. In addition, our proposal turns out to be an effective tool to deal with the portfolio concentration issues highlighted by \cite{BoCaPa}, mitigating the weight of the stock whose return represents the response variable of the estimated regression models. Finally,  the new strategies we propose in this study are computationally efficient, providing then an advantageous tool for several financial agents, such as traders and hedge funds, to improve the performance of their portfolios, while controlling the risk of incurring large losses during tail events.

Here, we build financial portfolios from quantile regression models estimated at the 5\% level. It would be interesting to analyse how the results change when using other quantile levels, not only the ones belonging to the left tail of the return distributions (reflecting a pessimistic asset allocation), but also those which refer to the right tail (suitable when markets are trending upward). Furthermore, we can also combine the strategies resulting from different quantile levels in the range $(0,1)$, adapting them to the different market conditions, exploiting the strengths of the different asset allocation schemes over time. We include these further analyses in our research agenda.

\bibliographystyle{Chicago}

\end{document}